\documentclass[11pt]{article}
\usepackage{moriond,epsfig,amsfonts}

\def\be{\begin{equation}}
\def\ee{\end{equation}}
\def\bea{\begin{eqnarray}}
\def\eea{\end{eqnarray}}

\begin{document}
\begin{flushright} 
UCRHEP-T389, May 2005
\end{flushright}
\vspace*{2cm}
\title{A NEW FAMILY SYMMETRY: DISCRETE QUATERNION GROUP
\footnote{Talk presented at the XL Rencontres de Moriond on Electroweak Interactions and Unified Theories, La Thuile, Italy, March 5-12, 2005.}}

\author{ M.~FRIGERIO }

\address{Department of Physics, University of California, Riverside, CA 92521, U.S.A.}

\maketitle\abstracts{
We examine the structure of the quaternion group $Q_8$ and its possible application to the physics of flavor. We find that a $Q_8$ family symmetry is suitable to explain the difference between quark and lepton mixing patterns. Distinctive phenomenological predictions are derived for the neutrino sector and the electroweak Higgs sector. We also show how the $Q_8$ symmetry suppresses the effective operators which mediate proton decay.
}

\section{Introduction}

The existence of non-zero neutrino masses, established in the last decade, requires some minimal extension of the Standard Model. If neutrinos are Majorana, present data can be described by a $3\times 3$ symmetric mass matrix ${\cal M}_\nu$, which adds 9 fundamental parameters to the 13 already present in the SM flavor sector.
A number of theoretical ideas developed in the attempt to understand the values of these parameters and to find an underlying symmetry principle.

In this talk we propose \cite{us} that the family symmetry may be related with a minimal discrete subgroup of $SU(2)$, the quaternion group $Q_8$. 
Its structure is suitable to accommodate three generations of quarks and leptons, in particular to explain the very large difference between the values of the $2-3$ mixing in the quark and lepton sectors. 
Different applications of discrete quaternion groups to flavor physics have been studied in the literature \cite{others}.

Some details of the relevant group theory are sketched in section \ref{group}.
The $Q_8$ model for fermion mixing is constructed in section \ref{model}.
In section \ref{pheno} we discuss the phenomenological predictions for neutrinos parameters, for the electroweak Higgs sector and for proton decay.

\section{Elements of quaternion group theory \label{group}}

A quaternion number can be written as $ q = a + i_1 b + i_2 c + i_3 d$, where 
$a,~b,~c,~d$ are real and $i_j$ are defined by $i_j^2=-1$ and $i_j i_k = \epsilon_{jkl} i_l$, that is, three imaginary units which do not commute. The set of quaternion numbers $Q$ is a group with respect to multiplication.
The quaternions $q$ with unit norm, $|q|\equiv\sqrt{a^2+b^2+c^2+d^2}=1$, form an invariant subgroup of $Q$ which is isomorphic to $SU(2)$.
The smallest non-trivial subgroup of $Q$ (non-trivial meaning that it is not a subgroup of the complex numbers) is known as $Q_8$. It is formed by the following 8 elements:
$\pm 1,~\pm i_1,~\pm i_2,~\pm i_3$. Since all these elements have norm 1, $Q_8$ is also a subgroup of $SU(2)$. A  faithful 2-dimensional representation is provided by the 8  special unitary $2\times 2$ matrices $\pm {\mathbb I}_2,~\pm i\sigma_{2}$, $\pm i\sigma_{1}$, $\pm i\sigma_{3}$, where ${\mathbb I}_2$ is the identity matrix and $\sigma_{i}$ are the Pauli matrices. 
Geometrically, $SU(2)$ is isomorphic to the 4-dimensional hyper-sphere ${\cal S}_3$ and $Q_8$ is the subgroup formed by the 8 vertices of the hyper-octahedron inscribed in ${\cal S}_3$ (the hyper-octahedron is the 4-dimensional perfect solid composed of 16 tetrahedra, also known as 16-cell).

The 8 elements of $Q_8$ are divided in 5 conjugacy classes, ${\cal C}_e = \{1\}$, ${\cal C}_{-e} = \{-1\}$, ${\cal C}_j = \{\pm i_j \}$, therefore there are 5 irreducible representations (irreps), which we denote ${\bf 1^{++}},~{\bf 1^{+-}},~{\bf 1^{-+}},~{\bf 1^{--}}$ and ${\bf 2}$.
These irreps derive from the decomposition of $SU(2)$ representations as follows:
${\bf 1}_{SU(2)} = {\bf 1^{++}}$, ${\bf 2}_{SU(2)} = {\bf 2}$, ${\bf 3}_{SU(2)} = 
{\bf 1^{+-}}+{\bf 1^{-+}}+{\bf 1^{--}}$. 
The characters can be easily constructed and are given in Table \ref{char}. The irreps ${\bf 1^{+-}},~{\bf 1^{-+}},~{\bf 1^{--}}$ share the same group properties and are therefore equivalent. This means that if a theory with $Q_8$ symmetry contains a set of these irreps, any permutation of them will not change the physical predictions.
The four 1-dimensional irreps combine as the irreps of a $Z_2 \times Z_2$ group:
${\bf 1^{s_1s_2}} \times {\bf 1^{s'_1s'_2}} = {\bf 1^{(s_1\cdot s'_1)(s_2\cdot s'_2)}}$.
The only non-trivial tensor product rule is ${\bf 2} \times {\bf 2} = {\bf 1^{++}} + {\bf 1^{+-}} + {\bf 1^{-+}} + {\bf 1^{--}}$, where $(\phi_1~\phi_2)^T, (\psi_1~\psi_2)^T \in {\bf 2}$ implies
\begin{equation}\begin{array}{ll}
(\phi_1\psi_2-\phi_2\psi_1) \in {\bf 1^{++}}~,~~~~~ & 
(\phi_1\psi_1 - \phi_2\psi_2) \in {\bf 1^{+-}}~,  \\
(\phi_1\psi_2+\phi_2\psi_1) \in {\bf 1^{-+}}~,~~~~~ & 
(\phi_1\psi_1 + \phi_2\psi_2) \in {\bf 1^{--}}~. \\
\end{array}\label{q8de}\end{equation}
As for $SU(2)$, $(\phi_1~\phi_2)^T \in {\bf 2}$ implies $(\phi_2^*~ - \phi_1^*)^T \in {\bf 2}$ (the ${\bf 2}$ irrep is complex but equivalent to its conjugate).

\begin{table}[t]
\caption{Character table of $Q_8$. Here $n$ is the number of 
elements in each conjugacy class, while $h$ is the order of any element $g$ 
in that class, i.e. the smallest integer such that $g^h = 1$.
}
\vspace{0.4cm}
\begin{center}
\begin{tabular}{|c|c|c|c|r|r|r|r|}
\hline
class & $n$ & $h$ & $\chi_{++}$ & $\chi_{+-}$ & $\chi_{-+}$ & 
$\chi_{--}$ & $\chi_{2}$ \\ 
\hline
${\cal C}_e$    & 1 & 1 & 1 & 1 & 1 & 1 & 2 \\ 
${\cal C}_{-e}$& 1 & 2 & 1 & 1 & 1 & 1 & $-2$ \\ 
${\cal C}_1$    & 2 & 4 & 1 & $-1$ & $-1$ & 1 & 0 \\ 
${\cal C}_2$    & 2 & 4 & 1 & 1 & $-1$ & $-1$ & 0 \\ 
${\cal C}_3$    & 2 & 4 & 1 & $-1$ & 1 & $-1$ & 0 \\
\hline
\end{tabular}
\end{center}
\label{char}\end{table}

\section{The fermion mixing in the presence of quaternion symmetry \label{model}}

\begin{figure}[t]\begin{center}
\psfig{figure=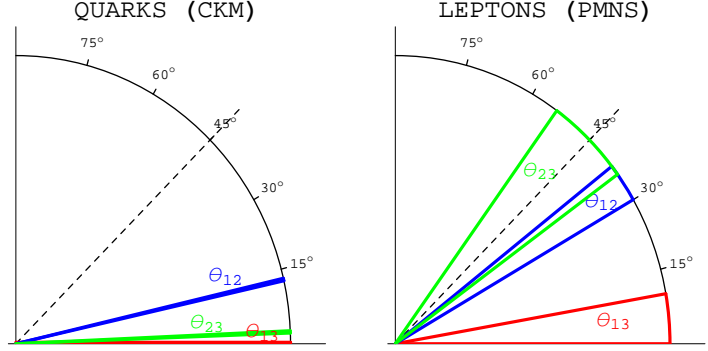,height=1.9in}
\caption{The fermion mixing angles in the quark and lepton sectors. 
The tiny quark $1-3$ mixing is non-zero, while only an upper bound is known for the lepton $1-3$ mixing.
\label{angles}}
\end{center}\end{figure}

In order to be guided in the search for a family symmetry, let us give a look to the values of mixing angles in the quark and lepton sectors, shown in
Fig.~\ref{angles} (the $99\%$ C.L. ranges for lepton mixing angles are taken from a global fit of neutrino oscillation data \cite{SV}). The prominent difference between CKM and PMNS mixing matrices appears in the $2-3$ sector: the mixing between 2nd and 3rd generation quarks is tiny, whereas muon and tau neutrinos mix almost maximally.

The determination of the mixing parameter in atmospheric neutrino experiments reached by now a significant precision, $\sin^2 2\theta^l_{23} \ge 0.91$ at $99\%$ C.L.. However this translates into a quite wide range for the leptonic $2-3$ mixing angle: $36^\circ \le \theta^l_{23} \le 54^\circ$. The most promising proposal to reduce this uncertainty significantly in the near future is probably the T2K experiment \cite{T2K}. Even though the deviation from maximal mixing may still be sizable,
there are several theoretical reasons to think that the value $\theta^l_{23}\sim\pi/4$ is not accidental and should be generated by some specific mechanism: 
(i) $\theta_{23}^l$ determines in most cases the dominant structure of the Majorana neutrino mass matrix ${\cal M}_\nu$;
(ii) radiative corrections from the superheavy scale (seesaw, GUT) to the electroweak scale do not generate naturally a large value of $\theta^l_{23}$ from a small one;
(iii) since $(\nu_\alpha ~ l_\alpha)^T$ is an $SU(2)_L$ doublet, $\nu_\alpha$ and $l_\alpha$ transform in the same way under any possible family symmetry, so that a cancellation is expected between the mixing in ${\cal M}_\nu$ and in the charged lepton mass matrix ${\cal M}_l$ ("flavor alignment"). 
This is the case for quarks: $\theta_{23}^q \approx 2^\circ$.

Let us show that the $Q_8$ family symmetry is appropriate to accommodate   large $\theta_{23}^l$ mixing and to reproduce the other main features of fermion mixing. In particular, $Q_8$ allows to distinguish the quark and lepton  2-3 sectors, assigning the 3 quark and lepton families as follows:
\begin{equation}
(u_i~d_i), ~u^c_i, ~d^c_i ~\in~ {\bf 1^{--}}, ~{\bf 1^{-+}}, 
~{\bf 1^{+-}}~; ~~~~~~~ 
(\nu_i~l_i), ~l^c_i ~\in~ {\bf 1^{++}}, ~{\bf 2}~.
\label{qrep}
\end{equation}
Electroweak symmetry breaking generates the mass terms ${\cal M}_f^{ij}f_i f^c_j$, $f=u,d,l$ and ${\cal M}_\nu^{ij}\nu_i\nu_j$. The matrix entries  are associated with the $Q_8$ assignment of the corresponding fermion bilinears: 
\begin{equation}
{\cal M}_{u,d} \sim \left(\begin{array}{ccc}
{\mathbf 1^{++}} & {\mathbf 1^{+-}} & {\mathbf 1^{-+}} \\
{\mathbf 1^{+-}} & {\mathbf 1^{++}} & {\mathbf 1^{--}} \\
{\mathbf 1^{-+}} & {\mathbf 1^{--}} & {\mathbf 1^{++}}
\end{array}\right)~,~~~~~
{\cal M}_{l,\nu} \sim \left(\begin{array}{ccc}
{\mathbf 1^{++}}& {\mathbf 2}			        & {\mathbf 2} \\
{\mathbf 2}        & {\mathbf 1^{--}}+{\mathbf 1^{+-}}& {\mathbf 1^{-+}}+{\mathbf 1^{++}} \\
{\mathbf 2}        & {\mathbf 1^{-+}}-{\mathbf 1^{++}}& {\mathbf 1^{--}}-{\mathbf 1^{+-}}
\end{array}\right)~.
\label{pattern}\end{equation}
(Since the Majorana matrix ${\cal M}_\nu$ is symmetric, the ${\bf 1^{++}}$ contribution in the $2-3$ neutrino sector is forbidden: $\nu_2\nu_3 - \nu_3\nu_2 = 0$.)
A non-zero value for a given entry of ${\cal M}_{u,d,l}$ (${\cal M}_\nu$) may be generated by the vacuum expectation value (VEV) of Higgs doublets $\phi_i$ (triplets $\xi_i$) transforming in the corresponding $Q_8$ irrep.
The triplet VEVs are induced \cite{masa} via trilinear couplings of the form $\mu_{ijk}\xi_i\phi_j\phi_k$ and are naturally tiny as long as the triplet masses are superheavy (type II seesaw mechanism). The scalar potential terms $\xi_i\phi_j\phi_k$ can be allowed to break the $Q_8$ symmetry softly, so that the $\xi_i$ VEVs are not constrained by the $Q_8$ assignments of $\phi_i$. Actually, it turns out that this $Q_8$ soft-breaking is necessary to generate a phenomenologically acceptable ${\cal M}_\nu$.

Taking into account Eq.~(\ref{pattern}), our construction proceeds as follows. Two Higgs doublets $(\phi_1^0 ~ \phi_1^-)^T \in {\bf 1^{++}}$ and
$(\phi_2^0 ~ \phi_2^-)^T \in {\bf 1^{+-}}$ generate 
\begin{equation}
{\cal M}_q = \left(\begin{array}{ccc} a_q & d_q & 0 \\ e_q & b_q & 0 \\ 0 & 0 & c_q
\end{array}\right)~,~~~~~~~~
{\cal M}_l = \left(\begin{array}{ccc} a_l & 0 & 0 \\ 0 & c_l & b_l \\ 0 & -b_l & -c_l
\end{array}\right)~,
\label{Mql}
\end{equation}
where $q=u,d$. 
The third quark families do not 
mix with the other two, i.e. $\theta^q_{13} = \theta^q_{23} = 0$, which 
is a good first approximation.  On the other hand, the ${\cal M}_l$ $2-3$ block is diagonalized by a 
rotation of $\pi/4$ on the left and on the right, i.e. $\mu,\tau = (l_2 
\pm l_3)/\sqrt 2$ and $\mu^c,\tau^c = (l_2^c \mp l_3^c)/\sqrt 2$.

At this point we have to choose the Higgs triplets $\xi_i$ appropriate to reproduce neutrino phenomenology. It turns out that the minimal number of triplets is four.
In particular, Eq.~(\ref{pattern}) shows that a non-zero neutrino $1-2$ mixing requires the VEVs of $(\xi_3,\xi_4)\in{\bf 2}$. In addition, one needs $\xi_1$ and $\xi_2$  transforming in two different  1-dimensional irreps. 
Four nonequivalent choices are possible, depending on the $Q_8$ assignments relative to $\phi_1 \in {\bf 1^{++}}$ and $\phi_2 \in {\bf 1^{+-}}$ (we remind that  ${\bf 1^{+-}},~{\bf 1^{-+}},~{\bf 1^{--}}$ are equivalent irreps, see section \ref{group}):
\begin{equation}\begin{array}{ll}
(1)~~ \xi_1 \in {\bf 1^{++}}~,~ \xi_2  \in {\bf 1^{+-}}~, &
(2)~~ \xi_1 \in {\bf 1^{++}}~,~ \xi_2  \in {\bf 1^{-+}} {\rm ~or~} {\bf 1^{--}} ~, \\ \\
(3)~~ \xi_1 \in {\bf 1^{-+}} {\rm ~or~} {\bf 1^{--}}~,~ \xi_2  \in {\bf 1^{+-}}~,~~~~~~~~ &
(4)~~ \xi_1 \in {\bf 1^{-+}}~,~ \xi_2  \in {\bf 1^{--}}~. 
\end{array}\end{equation}
In the basis where ${\cal M}_l$ is diagonal, the neutrino mass matrix ${\cal M}_\nu$
takes the following forms:
\begin{equation}\begin{array}{l}
(1)~~{\cal M}^{(e,\mu,\tau)}_\nu = \left(\begin{array}{ccc} a & c & d \\ c & 0 & b 
\\ d & b & 0\end{array}\right)~,~~~~~~~~~~~
(2)~~{\cal M}^{(e,\mu,\tau)}_\nu = \left(\begin{array}{ccc} a & c & d \\ c & b 
& 0 \\ d & 0 & b\end{array}\right)~,\label{one-two}\nonumber\\
(3)~~{\cal M}^{(e,\mu,\tau)}_\nu = \left(\begin{array}{ccc} 0 & c & d \\ c & a & b 
\\ d & b & a \end{array}\right)~,~~~~~~~~~~~
(4)~~{\cal M}^{(e,\mu,\tau)}_\nu = \left(\begin{array}{ccc} 0 & c & d \\ c & a 
& 0 \\ d & 0 & b\end{array}\right)~.\label{three-four}\nonumber
\end{array}
\label{mnus}\end{equation}
In all these scenarios ${\cal M}^{(e,\mu,\tau)}_\nu$ depends only on four parameters.
Since the form of ${\cal M}^{(e,\mu,\tau)}_\nu$ is a physical observable, it is invariant under a permutation of the ${\bf 1^{+-}}$, ${\bf 1^{-+}}$, ${\bf 1^{--}}$ assignments of quarks and Higgs bosons of the model, as one can check explicitly.

\section{Phenomenology of the $Q_8$ model \label{pheno}}

\subsection{Neutrinos} 

Let us discuss the predictions for neutrino phenomenology in the four scenarios of Eq.~(\ref{mnus}).

\begin{figure}[t!]\begin{center}
\psfig{figure=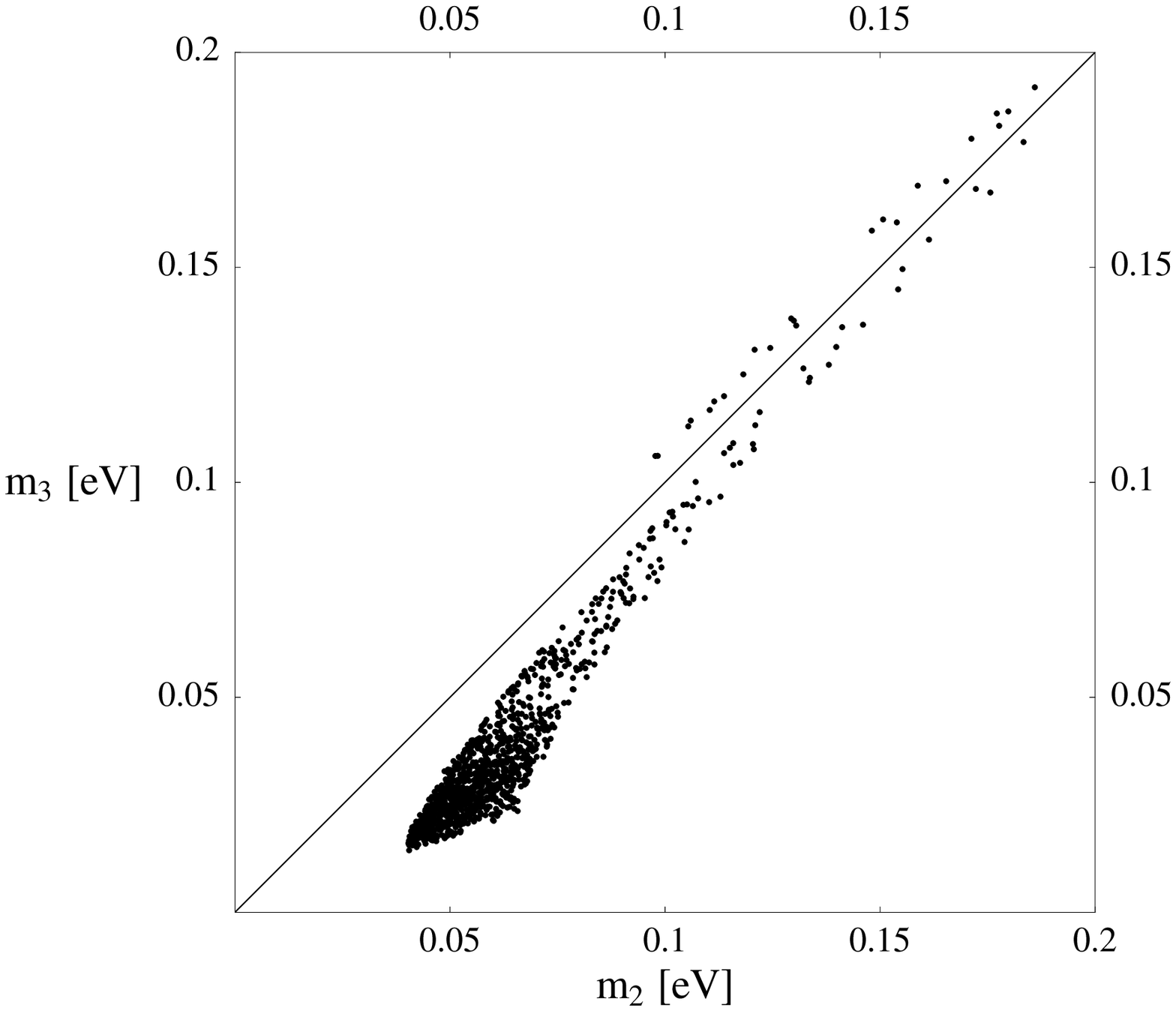,width=170pt}
~~~~~~~~
\psfig{figure=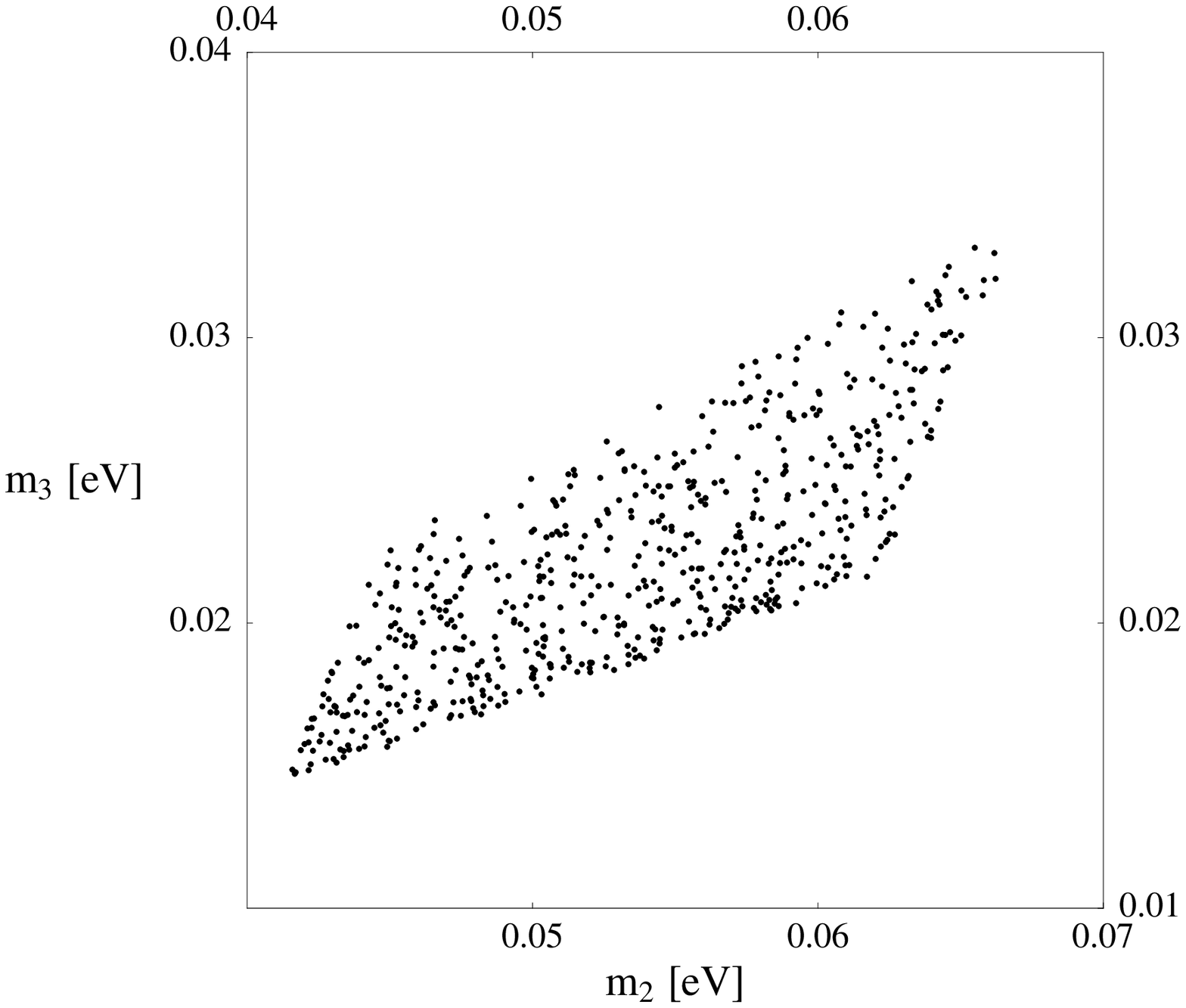,width=170pt}
\caption{In the left panel, the allowed region in $m_2 - m_3$ plane for scenario (1) is 
presented (the result for scenario (2) is very similar). 
The masses are scanned in the experimental allowed 
range : $\Delta m_{12}^2=(7.7 - 8.8)\times 10^{-5}$ eV$^2$, 
$\Delta m_{23}^2=(1.5 - 3.4)\times 10^{-3}$ eV$^2$, 
$\tan^2 \theta^l_{12}=0.33 - 0.49$, $\sin^2 2\theta^l_{23}\geq 0.92$, 
and $\sin\theta^l_{13}<0.2$.
In the right panel, the allowed region in $m_2 - m_3$ plane is shown for scenario (2) with the further assumption $\theta_{13}^l=0$.
\label{fignum}}
\end{center}\end{figure}

Scenario (1). Two texture zeros in the $\nu_\mu \nu_\mu$ and $\nu_\tau \nu_\tau$ 
entries are predicted by the $Q_8$ symmetry (the same matrix structure has been recently obtained \cite{GL} in a model with $Z_4$ family symmetry).
Here $\theta^l_{23} = \pi/4$ and $\theta^l_{13} = 0$ are obtained in the limit $c = \pm d$.  
Deviations from maximal $2-3$ mixing
as well as nonzero values of $\theta_{13}^l$ are allowed and  can be 
as large as the experimental upper bounds.
With the present experimental constraints, we find an inverted ordering of the mass spectrum with $|m_2| > 0.04$ eV and $|m_3| > 0.015$ eV, as shown in Fig.~\ref{fignum}, left panel.
The neutrinoless $2\beta$-decay rate is controlled by 
$m_{ee}\equiv |a| > 0.02$ eV.
A quasi-degenerate spectrum can be obtained, when 
$a\approx b$ and $c,d$ are much smaller; 
in this limit the ordering of the
spectrum can be also normal (see Fig.~\ref{fignum}, left panel).

Scenario (2).
In this case the $Q_8$ symmetry predicts one texture zero ($\mu\tau$ entry) and one equality of two matrix elements ($\mu\mu$ and $\tau\tau$). 
The phenomenology is in very good approximation the same as for scenario (1):
even though ${\cal M}^{(e,\mu,\tau)}_\nu$ seems very different in the two cases,
in the limit $\theta^l_{23} = \pi/4$ and $\theta^l_{13} = 0$ they
are distinguished only by the relative Majorana phase between $m_2$ and $m_3$, which is $-1$ for scenario (1) and to $+1$ for scenario (2). This phase is the unique physical parameter in ${\cal M}^{(e,\mu,\tau)}_\nu$ which cannot be measured with presently foreseeable techniques.  
In Fig.~\ref{fignum}, right panel, the allowed values of $m_{2,3}$ are shown in the limit $\theta_{13}^l=0$.

Scenario (3).
Also in this case the $Q_8$ symmetry predicts one texture zero ($ee$ entry) and one equality of two matrix elements ($\mu\mu$ and $\tau\tau$).
This structure implies a normal hierarchy of the mass spectrum. Here $\theta^l_{23} = \pi/4$ and $\theta^l_{13} = 0$ are again obtained in the limit $c=\pm d$. 
The constraint $\sin \theta^l_{13} < 0.2$ 
requires $\sin^2 2 \theta^l_{23} > 0.987$ and predicts
$0.035~{\rm eV} < |m_3| < 0.065~{\rm eV}$.
The neutrinoless $2\beta$ decay is suppressed, since $m_{ee}\equiv |(M_\nu^{(e,\mu,\tau)})_{11}|=0$.

Scenario (4).
This case is not viable: the texture zero in the $ee$ entry requires a mass spectrum with normal hierarchy, which is incompatible with the other zero in the $\mu\tau$ entry.

A recent paper \cite{KST} analyses all possible neutrino mass matrices with one texture zero and one equality of two non-zero matrix elements (as in scenarios (2) and (3)).

\subsection{Electroweak Higgs sector}

The $Q_8$ model contains two Higgs doublets, $\phi_1$ and $\phi_2$, distinguished by an odd-even parity.  The corresponding $Z_2$-symmetric scalar potential is well-known, having in general 
a minimum with nonzero VEVs for both $\phi_1^0$ and $\phi_2^0$, no CP violation and
all the 5 physical Higgs boson masses of the order of the electroweak scale.
However, the $Q_8$ ($Z_2$) symmetry may be softly broken by the term $m_{12}^2\phi_1^\dag \phi_2 +$ h.c., then CP is no longer conserved and the scalar masses can increase. More precisely, if $|m_{12}^2| \gg v^2$ one Higgs doublet decouples acquiring heavy mass $\sim |m_{12}^2|$, even though one can keep both $v_1,~v_2 \le v$. Nonetheless, it is reasonable to expect the soft-breaking scale to be smaller than or close to the electroweak scale $v$. In this case the non standard Higgs bosons may be light enough to produce some clear experimental signatures.

The off-diagonal couplings of $\phi_2$ to quarks in the $1-2$ sector (see Eq.~(\ref{Mql})) induce flavor changing neutral currents.
In particular, the non-standard neutral Higgs $h^0 = (v_1 \phi_2^0 - v_2 \phi_1^0)/\sqrt{v_1^2+v_2^2}$
contributes to the $K_L-K_S$ mass difference \cite{us}:
\begin{equation}
\frac{\Delta m_K}{m_K} \simeq \frac{B_K f_K^2}{3 m_h^2} 
\left( \frac{v_1^2+v_2^2} 
{v_1^2 v_2^2} \right) \sin^2 \theta_L \cos^2 \theta_L m_d m_s ~,
\end{equation}
where $m_h$ is the Higgs mass and $\theta_L$ is the $1-2$ left-handed mixing in ${\cal M}_d$.
Taking $v_1=v_2=123$ GeV, $\sin^2\theta_L \simeq m_d/m_s$, 
$B_K = 0.4$, $f_K = 114$ MeV and $m_d = 7$ MeV, this contribution 
is $1.1 \times 10^{-15} (100~{\rm GeV}/m_h)^2$, the 
experimental value being $7.0 \times 10^{-15}$.  
There are no flavor changing $\mu-\tau$ interactions because
${\cal M}_l$ in Eq.~(\ref{Mql}) is diagonalized by exactly maximal $2-3$ rotations. This implies that $h^0$ has the interaction
\begin{equation}
\frac{h^0}{2 \sqrt {v_1^2+v_2^2}} \left[ \left( \frac{v_1}{v_2} 
- \frac{v_2 }{v_1} \right) (m_\tau \tau \tau^c + m_\mu \mu \mu^c) +
\left( \frac{v_1}{v_2} + \frac{v_2}{v_1} \right) (m_\mu \tau \tau^c + 
m_\tau \mu \mu^c) \right] + h.c. ~.
\end{equation}
Therefore, the non-standard neutral Higgs mass eigenstates decay into $\tau^+ \tau^-$ and $\mu^+ \mu^-$ pairs with comparable strength ($\sim m_\tau/v$). This prediction may provide a crucial test of our model.

\subsection{Proton decay}

Let us discuss the implications of the $Q_8$ family symmetry for proton decay. Since no specific $B$-violating interaction is assumed in our model, we confront this issue in terms of the usual dimension 6 effective operators $qqql$, where $q$ denotes generically a quark field and $l$ a lepton. Let us remind that the action of the $Q_8$ symmetry does not depend on the specific chiralities of quarks and leptons (see Eq.~(\ref{qrep})). The unique $Q_8$ invariant operator is given by 
\begin{equation}
q_1 ~ q_2 ~ q_3 ~ l_1 ~,
\label{opp}\end{equation}
where the subscripts are family indexes in the $Q_8$ symmetry basis.  Eq.~(\ref{Mql}) implies that $q_1$ and $q_2$ are mixtures of first and second generation quark mass eigenstates, $q_3$ is identified with the bottom or top quark, $l_1$ with the electron or the electron neutrino. As a consequence, the operator (\ref{opp}) cannot mediate proton decay, since it involves a third generation quark. 
However, the experimentally tiny but non-zero values of the $2-3$ and $1-3$ CKM mixing angles indicate that Eq.~(\ref{Mql})
is valid only in first approximation, that is in the limit where top/bottom flavor numbers are unbroken global symmetries.
They may be broken, for example, by adding an Higgs doublet $\phi_3\in {\bf 1^{--}}$, which contributes to the (23) and (32) entries of ${\cal M}_{u,d}$.
In a realistic case, therefore, the operator (\ref{opp}) may contribute to proton decay, but only through the very small mixing angle $\theta_{13}^{CKM}$ (or $\theta_{23}^{CKM}$ for decays into strange mesons). Therefore one expects an enhancement of the proton lifetime with respect to models with generic dimension 6 operators, but the actual prediction requires to specify the scale and nature of B-violating interactions.

\section{Conclusions}

We disclosed the role of the discrete quaternion group $Q_8$ in understanding fermion mixing. Our model predicts three possible structures for the Majorana neutrino mass matrix, which accommodate present data and can be ruled out by future experiments. The model predicts also non-standard Higgs bosons decaying into $\mu^+\mu^-$ and $\tau^+\tau^-$ with comparable rates. Finally, the $Q_8$ symmetry partially suppresses  the effective operators which mediate proton decay.

\section*{Acknowledgments}
This talk substantially relies on the work done in collaboration with S.Kaneko, E.Ma and M.Tanimoto \cite{us}. I thank G.Senjanovic and J.-M.Frere for asking useful questions and C.Mariani and E.Ma for fruitful discussions. My participation to the Moriond Conference was supported by the European Union Program "Human Resources and Mobility Activity - Marie Curie Conferences" and by the U.S. Department of Energy under Grant No. DE-FG03-94ER40837.

\section*{References}

\end{document}